\documentclass[12pt]{article}

\newcommand\beq{\begin{equation}}
\newcommand\eeq{\end{equation}}

\date{}
\usepackage[final]{graphicx}
\usepackage{amsmath}
\usepackage{amsfonts}   
\usepackage{amssymb}
\begin{document}


\title{Parameterizing Quasiperiodicity: Generalized Poisson Summation and Its Application to Modified-Fibonacci Antenna Arrays}

\author{Vincenzo Galdi, Giuseppe Castaldi, 
Vincenzo Pierro,\\ Innocenzo M. Pinto, and
Leopold B. Felsen 
\thanks{
V. Galdi, G. Castaldi, V. Pierro and I.M. Pinto are with the Waves Group, Department of
Engineering, University of Sannio, I-82100 Benevento, Italy (e-mail: vgaldi@unisannio.it, castaldi@unisannio.it, pierro@unisannio.it, pinto@sa.infn.it).\protect\\
L.B. Felsen is with the Department of Aerospace and Mechanical Engineering and the
Department of Electrical and Computer Engineering (part-time), Boston University, Boston, MA 02215 USA. He is also University Professor Emeritus, Polytechnic University, Brooklyn,
NY 11201 USA (e-mail: lfelsen@bu.edu).}}

\maketitle

\begin{abstract}
The fairly recent discovery of ``quasicrystals'', whose X-ray diffraction patterns reveal certain peculiar features which do not conform with spatial periodicity, has motivated studies of the wave-dynamical implications of ``aperiodic order''. 
Within the context of the radiation properties of antenna arrays, an instructive novel (canonical)
example of wave interactions with {\em quasiperiodic} order is illustrated here for one-dimensional (1-D) array configurations based on the ``modified-Fibonacci'' sequence, with utilization of a two-scale generalization of the standard Poisson summation formula for periodic arrays.
This allows for a ``quasi-Floquet'' analytic parameterization of the radiated field, which provides instructive insights into some of the basic wave mechanisms associated with quasiperiodic order, highlighting similarities and differences with the periodic case.
Examples are shown for quasiperiodic infinite and spatially-truncated arrays, with brief discussion of 
computational issues and
potential applications.
\end{abstract}

\markboth{GALDI et al.: Parameterizing Quasiperiodicity: Generalized Poisson Summation...}{} 

\section{Introduction}
\label{intro}
The recent discovery (1984) of ``quasicrystals'', i.e., certain metallic alloys whose X-ray diffraction patterns contain bright spots displaying symmetries (e.g., 5-fold) which are {\em incompatible} with {\em spatial periodicity} \cite{Shechtman}, \cite{Levine}, has stimulated a growing interest in the study of {\em aperiodic order} and its wave-dynamical properties. 

In electromagnetics (EM) engineering, use of random or deterministic {\em aperiodic} geometries has been customary within the framework of antenna array {\em thinning} \cite{Mailloux}--\cite{fractarr1}, whereas {\em multiperiod} configurations have recently been proposed for optimizing the passband/stopband characteristics of frequency selective surfaces \cite{Mittra} and photonic bandgap (PBG) devices \cite{Parini}. In \cite{Tiling_AP}, we explored the radiation properties of two-dimensional (2-D) antenna arrays based on the concept of ``aperiodic tiling'' \cite{Grunbaum}, \cite{Senechal1}, which had previously found interesting applications in the field of PBG devices (see the brief summary and references in \cite{Tiling_AP}).

In this paper, we turn our attention to the radiation properties of a different category of aperiodic structures consisting of two-scale Fibonacci-type sequences \cite{Fib1}. Materials exhibiting this type of aperiodic order, technically called ``quasiperiodicity'' (see the definition in Sec. \ref{Mod_Fib}), were first fabricated in 1985 as GaAs-AlAs heterostructures (1-D multilayers) \cite{Merlin}; their technical classification as ``quasicrystals'' is still debated \cite{Lifshitz}. The wave-dynamical properties of 1-D and 2-D Fibonacci-type structures have been widely investigated, theoretically and experimentally, in quantum mechanics, acoustics and EM (see \cite{Kohmoto}--\cite{Lifshitz1} for a sparse sampling). Particularly interesting outcomes concern the self-similar fractal (Cantor-type) nature of the eigenspectra \cite{Kohmoto}, \cite{Suto}, \cite{Vasco}, and the possible presence of bandgaps \cite{Macia}, \cite{Ouyang}, omnidirectional reflection properties \cite{Cojo}, \cite{Dong}, and  localization phenomena \cite{Kohmoto}, \cite{Local}, \cite{DalNegro}.

Here, we concentrate on the study of the radiation properties of a simple class of 1-D antenna arrays based on the so-called ``modified-Fibonacci'' sequence. This novel prototype array configuration appears particularly well suited to exploration of some basic characteristics of wave interactions with quasiperiodic order.
First, this array model is amenable to {\em analytic} parameterization via a {\em generalized Poisson summation} formula, by exploiting some recent results in \cite{Buczek} (this paper is easily accessible through the web). Note that, in principle, one can derive generalized Poisson summation formulas that accommodate a variety of ``nonperiodicity scales'' (see, e.g., \cite{Chen}). Here, we consider only two scales, $d_1$ and $d_2$.
This opens up the possibility of extending the Floquet-based parameterization of infinite and semi-infinite time-harmonic {\em periodic} arrays in \cite{Capolino1}, \cite{Capolino2} to the case of two-scale {\em quasiperiodic} arrays. Next, the inherent degree of freedom in the choice of the ratio between the two scales can be used to study the ``transition'' from periodic ($d_1=d_2$) to quasiperiodic ($d_1\ne d_2$) order, so as to better understand the quasiperiodicity-induced footprints in the wave dynamics. Finally, although the main focus of this preliminary investigation is on wave-dynamical phenomenologies, computational and applicational issues for a test example are briefly addressed as well, including possible exploitation of the two-scale degree of freedom for pattern control in practical applications.

The paper is organized as follows. In Section \ref{formulation}, the problem geometry is described, and the modified-Fibonacci sequence is introduced together with general aspects of quasiperiodicity. In Section \ref{Gen_Poiss}, the generalized Poisson summation formula for modified-Fibonacci arrays is introduced, and its similarities and differences with the periodic case are discussed. In Section \ref{Radiation}, a ``quasi-Floquet'' (QF) parameterization of the radiated field for infinite and semi-infinite modified-Fibonacci arrays is derived, paralleling \cite{Capolino1}, \cite{Capolino2}. Numerical results and potential applications are illustrated in Section \ref{Res_App}, followed by brief concluding remarks in Section \ref{Conclusions}.

\section{Background and Problem Formulation}
\label{formulation}

\subsection{Geometry}
\label{Geom}
\begin{figure} 
\begin{center}
\includegraphics[width=8cm]{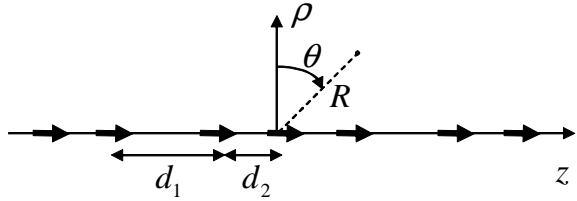} 
\end{center}
\caption{Problem schematic: An infinite (or semi-infinite) phased line array of $z$-directed electric dipoles is considered. The dipole distribution $z_m$, which features only two possible inter-element spacings $d_1$ and $d_2\le d_1$, is chosen according to the modified-Fibonacci sequence in (\ref{eq:ModFib}). Also shown are the $(z,\rho)$ and $(R,\theta)$ coordinate systems utilized.} 
\label{Figure1} 
\end{figure}

Referring to the geometry depicted in Fig. \ref{Figure1}, we begin with an infinite phased line array of $z$-directed electric dipoles, subject to uniform unit-amplitude time-harmonic $\exp(j\omega t)$ excitation, described by the current distribution
\beq
f(z)=\sum_{m=-\infty}^{\infty}\delta(z-z_m) \exp\left(-j\eta k_0 z_m\right),
\label{eq:dist1}
\eeq
where $k_0=\omega\sqrt{\epsilon_0\mu_0}=2\pi/\lambda_0$ is the free-space wavenumber (with $\lambda_0$ being the wavelength), and $-1\le\eta\le1$ describes the inter-element phasing. The dipole sequence $\left\{z_m\right\}_{m=-\infty}^{\infty}$, which is restricted to two possible inter-element spacings $d_1$ and $d_2\le d_1$ (see Fig. \ref{Figure1}), is chosen according to the modified-Fibonacci rule, whose properties are summarized below. Spatial truncation effects will be discussed in Sec. \ref{Semi-Infinite}.

\subsection{Modified-Fibonacci Sequence: An Example of\\ Quasiperiodic Order}
\label{Mod_Fib}
The Fibonacci sequence \cite{Fib1}, introduced in 1202 by the Italian mathematician Leonardo da Pisa (Fibonacci) (ca. 1170--ca. 1240), in connection with a model for rabbit breeding, is probably the earliest and most thoroughly investigated deterministic aperiodic sequence. 
Since then, it has found applications in many different fields, owing to its intimate relation with one of the most pervasive mathematical entities, the {\em Golden Mean} \cite{Fib1}. 
In its simplest version, the Fibonacci sequence can be generated from a two-symbol alphabet ${\cal A}=\left\{a,b\right\}$, by iteratively applying the substitution rules
\beq
a\rightarrow ab,~~b\rightarrow a,
\eeq
so as to construct a sequence of symbolic strings $s_n$
\beq
b \Rightarrow a \Rightarrow ab
\Rightarrow aba
\Rightarrow abaab
\Rightarrow abaababa 
\Rightarrow abaababaabaab
\cdots
\label{eq:Fibstrings}
\eeq
Note that the string at each iteration is obtained as the concatenation of the two preceding ones ($s_n=s_{n-1} \cup s_{n-2}$).
The process can be iterated {\em ad infinitum}, yielding an infinite sequence of ``$a$'' and ``$b$'' symbols which seems
to display no apparent regularity, but actually hides a wealth of interesting properties (see \cite{Fib1} for details). 
For instance, in the limit of an infinite sequence, it can be shown that the ratio between the numbers of ``$a$'' and ``$b$'' symbols ($N_a$ and $N_b$, respectively) approaches the Golden Mean $\tau=(1+\sqrt{5})/2$ \cite{Fib1},
\beq
\lim_{N\rightarrow\infty}\frac{N_a}{N_b}=\tau=\frac{1+\sqrt{5}}{2},~~~N_a+N_b=N.
\label{eq:limit}
\eeq

In connection with the antenna array problem of interest here, there are several ways of embedding the above-introduced Fibonacci-type aperiodic sequence. One possibility would be keeping the geometry {\em periodic} (i.e., uniform inter-element spacing) and associating with the ``$a$'' and ``$b$'' symbols in the Fibonacci sequence two possible current amplitudes. Another possibility, pursued in this investigation, assumes {\em uniform excitation} and associates the ``$a$'' and ``$b$'' symbols with two possible inter-element spacings, $d_1$ and $d_2\le d_1$, respectively (see Fig. \ref{Figure1}). Accordingly, the dipole positions $z_m$ in (\ref{eq:dist1}) can be obtained from the symbolic strings in (\ref{eq:Fibstrings}) or, in an equivalent and more direct fashion, via \cite{Buczek}
\beq
z_m=d_1 \left\|\frac{m}{\tau}\right\|+d_2\left(m-\left\|\frac{m}{\tau}\right\|\right),
\label{eq:ModFib}
\eeq
where $\left\|\cdot\right\|$ denotes the nearest-integer function,
\beq
\left\|x\right\|=\left\{
\begin{array}{ll}
m,~~~~~~~~~~~~m\le x< m+\frac{1}{2},\\
m+1,~~~~~~m+\frac{1}{2}\le x\le m+1.
\end{array}
\right.
\eeq

Following \cite{Buczek}, the particular case $d_2=d_1/\tau$ in (\ref{eq:ModFib}) will be referred to as ``standard Fibonacci''. The general case, in which the scale ratio $\nu=d_2/d_1$ is left as a degree of freedom, will be referred to as ``modified Fibonacci'', and represents one of the simplest extensions of the Fibonacci sequence. The reader is referred to \cite{Chen}, \cite{Severin}, \cite{Wolny} for other examples of possible extensions/generalizations.

Note that the modified-Fibonacci sequence in (\ref{eq:ModFib}) includes as a limit the {\em periodic} case ($d_1=d_2$). It can be proved that, with the exception of this degenerate case, the sequence in (\ref{eq:ModFib}) and the corresponding dipole distribution in (\ref{eq:dist1}) display a ``quasiperiodic'' character, which represents one of the most common and best known forms of aperiodic order. The concept of quasiperiodicity stems from the theory of ``almost-periodic'' functions developed by H. Bohr during the 1920s \cite{quasip1}, \cite{quasip2}. In essence, a quasiperiodic function can be uniformly approximated by a generalized Fourier series containing a countable infinity of {\em pairwise incommensurate} frequencies generated from a finite-dimensional basis (see \cite{quasip1}, \cite{quasip2} for more details). 

It follows from (\ref{eq:limit}) that, in the infinite-sequence limit, the average inter-element spacing $d_{av}$ is \cite{Buczek}
\beq
d_{av}=\frac{\tau d_1+d_2}{1+\tau}.
\label{eq:dav}
\eeq
Anticipating the analytic derivations and parametric analysis in Secs. \ref{Gen_Poiss} and \ref{Radiation},
it is expedient to parameterize the sequence in (\ref{eq:ModFib}) in terms of the average spacing $d_{av}$ in (\ref{eq:dav}) and the scale ratio $\nu=d_2/d_1$, rewriting $d_1$ and $d_2$ as
\beq
d_1=\frac{(1+\tau)}{(\nu+\tau)}d_{av},~~~
d_2=\nu d_1,~~0< \nu\le 1.
\eeq

\begin{figure} 
\begin{center}
\includegraphics[width=12cm]{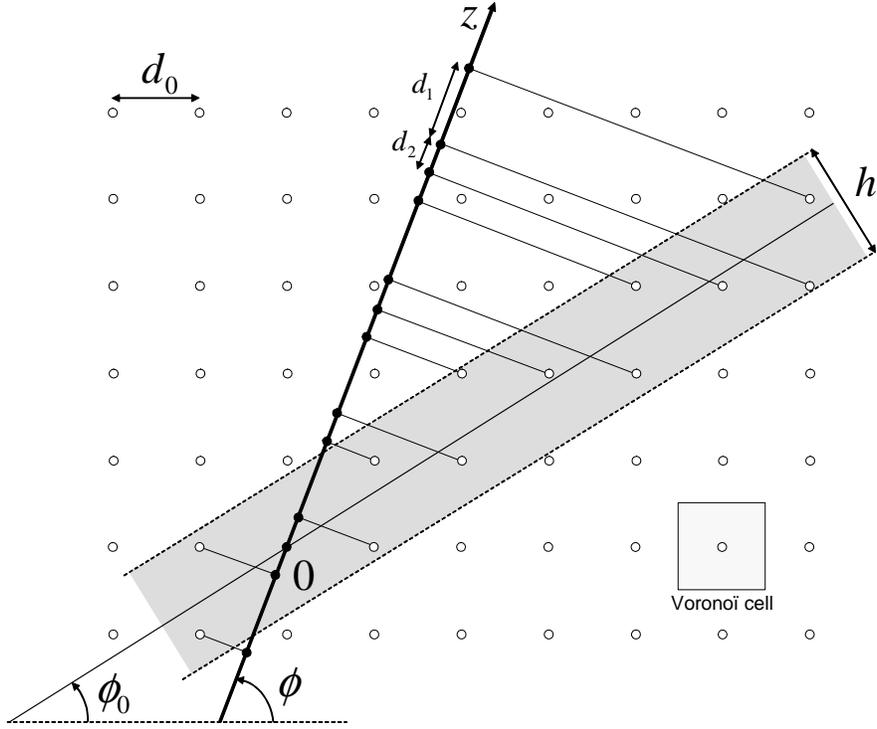} 
\end{center}
\caption{``Cut-and-project'' graphic interpretation of the modified-Fibonacci sequence in (\ref{eq:ModFib}). A 2-D square lattice of side $d_0=d_1\sqrt{1+\nu^2}$ is traversed by a straight line with slope $\tan\phi_0=1/\tau$, and those lattice points whose ``Vorono\"i cell'' (light-shaded square cell of side $d_0$ centered around the point) is crossed by the line (or, equivalently, those falling within the dark-shaded rectangular window of size $h=d_0(1+\tau)/\sqrt{2+\tau}$ centered around the line) are orthogonally projected onto another straight line ($z$-axis) with slope $\tan\phi=\nu$. For the standard-Fibonacci sequence ($\nu=1/\tau$), the two lines coincide  ($\phi=\phi_0$).} 
\label{Figure2} 
\end{figure}

The modified-Fibonacci sequence in (\ref{eq:ModFib}) admits an instructive alternative interpretation in terms of a ``cut-and-project'' graphic construction \cite{Buczek}, as illustrated in Fig. \ref{Figure2}.
Cut-and-project schemes are systematic tools for generating {\em quasiperiodic} sets via projection from higher-dimensional {\em periodic} lattices (see \cite{Senechal1} for details). In our case, one starts from a 2-D square lattice of side $d_0=d_1\sqrt{1+\nu^2}$ traversed by a straight line with slope $\tan\phi_0=1/\tau$. Those lattice points whose ``Vorono\"i cell'' \cite{Senechal1} (light-shaded square cell of side $d_0$ centered around the point, in Fig. \ref{Figure2}) is crossed by the line (or, equivalently, those falling within the rectangular window of size $h=d_0(1+\tau)/\sqrt{2+\tau}$ centered around the line, in Fig. \ref{Figure2}) are orthogonally projected onto another straight line ($z$-axis in Fig. \ref{Figure2}) with slope $\tan\phi=\nu$ to yield the desired modified-Fibonacci sequence in (\ref{eq:ModFib}). For the standard-Fibonacci sequence ($\nu=1/\tau$), the two lines coincide ($\phi=\phi_0$) and the above scheme becomes equivalent to the canonical cut-and-project scheme described in \cite{Senechal1}.

\section{Generalized Poisson Summation Formula}
\label{Gen_Poiss}
For {\em periodic} structures, Floquet theory provides a rigorous and powerful framework for spectral- or spatial-domain analytic and numerical analysis. In this connection, the Poisson summation formula \cite{Papoulis} can be utilized to systematically recast field observables as superpositions of either {\em individual} or {\em collective} contributions. Problem-matched extensions have also been developed to accommodate typical departures from perfect periodicity in realistic structures, such as truncation (finiteness) and {\em smooth} perturbations in the spatial period as well as in the excitation (tapering) \cite{Felsen1}--\cite{Maci}.

Considering the degenerate {\em periodic} limiting case ($d_1=d_2=d$, i.e., $z_m=md$) of the modified-Fibonacci sequence in (\ref{eq:ModFib}), the corresponding periodic limit $f^{(per)}$ of the current distribution in (\ref{eq:dist1}) can be recast, via the standard Poisson summation formula \cite{Papoulis}, as
\begin{subequations}
\begin{eqnarray}
f^{(per)}(z)=\sum_{m=-\infty}^{\infty}\delta(z-m d)\exp\left(-j\eta k_0 md\right)
\label{eq:Poiss_1}\\
=
\frac{1}{d}\sum_{q=-\infty}^{\infty}
\exp\left(-jk_{zq}z\right),~~
k_{zq}=k_0\eta+\frac{2\pi q}{d}.
\label{eq:Poiss2}
\end{eqnarray}
\label{eq:Poiss}
\end{subequations}
The $m$-indexed {\em individual} dipole contributions in (\ref{eq:Poiss_1}) are thereby recast into the infinite superposition of linearly smoothly-phased $q$-indexed equivalent line source distributions in (\ref{eq:Poiss2}). Remarkably, a similar re-parameterization is {\em always} possible for the {\em general} (i.e., {\em quasiperiodic}) case of the modified-Fibonacci sequence in (\ref{eq:ModFib}), since it can be shown that the spatial Fourier transform (plane-wave spectrum) of (\ref{eq:dist1}) can be written as \cite{Buczek} (see also the Appendix)
\begin{subequations}
\begin{eqnarray}
{\hat F}(k_z)&=&\int_{-\infty}^{\infty}f(z)\exp(j k_z z) dz\nonumber\\
&=&\sum_{m=-\infty}^{\infty}\exp(jk_z z_m)\exp\left(-j\eta k_0 z_m\right)
\label{eq:Fib_spect1}\\
&=&\frac{2\pi}{d_{av}}\sum_{q_1,q_2=-\infty}^{\infty}S_{q_1q_2}
\delta(k_z-k_{zq_1q_2}),
\label{eq:Fib_spect2}
\end{eqnarray}
\label{eq:Fib_spect}
\end{subequations}
where $d_{av}$ is defined in (\ref{eq:dav}), and the amplitude coefficients $S_{q_1q_2}$ and the spatial frequencies $k_{zq_1q_2}$ are given by 
\begin{eqnarray}
S_{q_1q_2}&=&\frac{\sin W_{q_1q_2}}{W_{q_1q_2}},\nonumber\\
W_{q_1q_2}&=&\frac{\pi}{d_{av}}\left(q_1 d_1 -q_2d_2\right)
=\frac{\pi(1+\tau)(q_1-q_2\nu)}{\nu+\tau},
\label{eq:Wq}
\end{eqnarray}
\beq
k_{zq_1q_2}=k_0\eta+\alpha_{q_1q_2},~~\alpha_{q_1q_2}=\frac{2\pi}{d_{av}}\frac{(q_1+q_2\tau)}{(\tau+1)}.
\label{eq:spat_freq}
\eeq
It then follows via straightforward inverse Fourier transform of (\ref{eq:Fib_spect2}) that
\beq
f(z)=
\frac{1}{d_{av}}\sum_{q_1,q_2=-\infty}^{\infty}S_{q_1q_2}
\exp\left(-jk_{zq_1q_2}z\right).
\label{eq:Gen_Poiss}
\eeq
Equation (\ref{eq:Gen_Poiss}) thus represents a generalization of the standard Poisson summation formula for {\em periodic} arrays in (\ref{eq:Poiss}) to the more general {\em quasiperiodic} modified-Fibonacci array in (\ref{eq:ModFib}), and will accordingly be referred to as {\em generalized} Poisson summation formula. It is readily verified that for the special case of {\em periodic} arrays ($d_1=d_2=d$, i.e., $\nu=1$), one obtains
\beq
d_{av}=d,~~W_{q_1q_2}=(q_1-q_2)\pi,~~S_{q_1q_2}=\delta_{q_1q_2}, 
\eeq
with $\delta_{q_1q_2}$ denoting the Kronecker delta, and thus (\ref{eq:Gen_Poiss}) reduces to the standard Poisson summation formula in (\ref{eq:Poiss2}).

A few observations are in order to highlight similarities and differences between (\ref{eq:Gen_Poiss}) and (\ref{eq:Poiss}). Similar to (\ref{eq:Poiss}), the {\em individual} dipole contributions in (\ref{eq:ModFib}) are recast via (\ref{eq:Gen_Poiss}) as a superposition of {\em collective} contributions arising from linearly smoothly-phased equivalent line source distributions. However, at variance with the $q$-indexed {\em single} infinity of {\em equispaced} spatial frequencies in (\ref{eq:Poiss2}), the generalized Poisson summation formula in (\ref{eq:Gen_Poiss}) entails a $(q_1,q_2)$-indexed {\em double} infinity of generally {\em pairwise-incommensurate} spatial frequencies. This could be expected recalling the definition of the quasiperiodic functions given in Sec. \ref{Mod_Fib}.
Note that the spatial frequencies $k_{zq_1q_2}$ in (\ref{eq:spat_freq}) depend on the average inter-element spacing $d_{av}$, but are {\em independent} of the scale ratio $\nu=d_2/d_1$. The dependence on $\nu$ shows up (via $W_{q_1q_2}$) in the amplitude coefficients $S_{q_1q_2}$ in (\ref{eq:Wq}), which are not constant as for the periodic case in (\ref{eq:Poiss2}). 
It is worth pointing out that such dependence is {\em smooth}, and consequently no {\em abrupt} transition occurs when $\nu$ is varied between {\em rational} and {\em irrational} values, i.e., going from {\em commensurate} to {\em incommensurate} scales $d_1$ and $d_2$. However, for case of {\em commensurate} scales, 
\beq
\nu=\frac{d_2}{d_1}=\frac{p_1}{p_2},~~p_1, p_2\in {\mathbb N},
\eeq
one can readily verify that
\begin{eqnarray}
k_{z(q_1+Mp_1)(q_2+Mp_2)}&=&k_{zq_1q_2}+M\left(\frac{2\pi}{d_1}p_2\right),\nonumber\\
W_{(q_1+Mp_1)(q_2+Mp_2)}&=&W_{q_1q_2},~~M\in{\mathbb Z},
\end{eqnarray}
and thus the array spectrum in (\ref{eq:Fib_spect2}) is {\em periodic} with period $2\pi p_2/d_1$. It can be verified that the opposite implication is also true, and thus the spectrum is periodic {\em if and only if} the scales $d_1$ and $d_2$ are commensurate, although the array remains {\em aperiodic} in physical space \cite{Buczek}.

\section{Radiated Field}
\label{Radiation}
In order to further explore possible similarities and differences between the wave phenomenologies associated with {\em periodicity} and {\em quasiperiodicity}, it is instructive to consider the radiated fields. To this end, in what follows, we extend the Floquet-based frequency-domain analysis in \cite{Capolino1}, \cite{Capolino2}, for infinite and semi-infinite periodic phased arrays of dipoles, to the quasiperiodic modified-Fibonacci case.
As in \cite{Capolino1}, \cite{Capolino2}, attention is restricted to the $z$-directed vector potential ${\bf A}({\bf r})=A({\bf r}){\bf u}_z$, with ${\bf r}\equiv(z,\rho)$ and with ${\bf u}_z$ denoting a $z$-directed unit vector, from which all field quantities of interest can be computed. Whenever applicable, partial results from \cite{Capolino1}, \cite{Capolino2} are recalled and used, without going into the details of the technical derivations.

\subsection{Infinite Arrays}
\label{Infinite}
For infinite arrays, proceding analogous to \cite{Capolino1}, the generalized Poisson summation formula in (\ref{eq:Gen_Poiss}) can be used to recast the element-by-element (spherical wave) synthesis of the field potential,
\begin{eqnarray}
A({\bf r})&=&\int_{-\infty}^{\infty}f(z')\frac{\exp\left(-jk_0R\right)}{4\pi R}dz'=
\sum_{m=-\infty}^{\infty}A_m({\bf r})\nonumber\\
&=&\sum_{m=-\infty}^{\infty}\frac{\exp\left(-jk_0R_m\right)}{4\pi R_m} \exp\left(-jk_0\eta z_m\right),
\label{eq:elbyel}
\end{eqnarray}
where $R=\sqrt{\rho^2+(z-z')^2}$ and $R_m=\sqrt{\rho^2+(z-z_m)^2}$, in terms of a ``quasi-Floquet'' (QF) representation
\beq
A({\bf r})=\sum_{q_1,q_2=-\infty}^{\infty}S_{q_1q_2} A^{QF}_{q_1q_2}({\bf r}),
\label{eq:QFS}
\eeq
\begin{eqnarray}
A^{QF}_{q_1q_2}({\bf r})&=&\frac{1}{d_{av}} \int_{-\infty}^{\infty}
\frac{\exp\left(-jk_0R\right)}{4\pi R}\exp\left(-jk_{zq_1q_2}z'\right)dz'\nonumber\\
&=&
\frac{\exp\left(-jk_{zq_1q_2}z\right)}{4j d_{av}}
\mbox{H}_0^{(2)}\left(k_{\rho q_1 q_2}\rho\right).
\label{eq:QFW1}
\end{eqnarray}
In (\ref{eq:QFW1}), $\mbox{H}_0^{(2)}$ denotes the zeroth-order Hankel function of the second kind (line-source Green's function), and $k_{\rho q_1 q_2}=\sqrt{k_0^2-k^2_{z q_1 q_2}}$, $\mbox{Im}\left(k_{\rho q_1 q_2}\right)\le 0$
denote the QF wave radial wavenumbers related to the $z$-domain wavenumbers $k_{zq_1q_2}$ in (\ref{eq:spat_freq}). 
Recalling the asymptotic ($|k_{\rho q_1 q_2}\rho|\rightarrow\infty$) expansion of the Hankel function, one obtains
\begin{eqnarray}
A^{QF}_{q_1q_2}({\bf r})&&\sim
\frac{1}{d_{av}\sqrt{4\pi k_{\rho q_1 q_2}\rho}}\nonumber\\
&\times &\exp\left[-j\left(k_{\rho q_1 q_2} \rho + k_{z q_1 q_2}z+\frac{\pi}{4}\right)\right],
\label{eq:PQFW}
\end{eqnarray}
from which it is recognized that
\beq
\left|k_{z q_1 q_2}\right|< k_0,~~~q_2\lessgtr-\frac{q_1}{\tau}\pm\left(\frac{d_{av}}{\lambda_0}\right)
\frac{\left(1\mp\eta\right)\left(1+\tau\right)}{\tau},
\label{eq:prop}
\eeq
corresponds to {\em radially-propagating} QF waves, whereas
$|k_{z q_1 q_2}|> k_0$ corresponds to {\em radially-evanescent} QF waves. Accordingly, sufficiently far from the array axis, the potential field in (\ref{eq:QFS}) will be synthesized in terms of {\em conical} propagating QF waves (heavy solid arrow in Fig. \ref{Figure3}), with arrival directions
\beq
\beta_{q_1q_2}=\arccos\left(\frac{k_{z q_1 q_2}}{k_0}\right).
\label{eq:betaq1q2}
\eeq
In the $(q_1,q_2)$-plane, the propagating spectral range in (\ref{eq:prop}) is mapped into an {\em infinite} strip, thus indicating the general presence of an {\em infinite} number of propagating QF waves. This is illustrated in Fig. \ref{Figure4}, for a nonphased ($\eta=0$) array with $d_{av}=1.1\lambda_0$ and various values of the scale ratio $\nu=d_2/d_1$. Also shown, for comparison, is the {\em periodic} case (Fig. \ref{Figure4}(a)), which entails a {\em finite} number of propagating Floquet waves. In the general {\em quasiperiodic} case, it is observed that the propagating spectral range is vastly populated, irrespective of the commensurate (Fig. \ref{Figure4}(b)) or incommensurate (Fig. \ref{Figure4}(c)) character of the scales $d_1$ and $d_2$. However, it follows from (\ref{eq:Wq}) that moving toward large values of $q_1, q_2$ (with opposite sign), the amplitude coefficients $S_{q_1q_2}$ decay non-monotonically as $\sim (q_1-q_2\nu)^{-1}$. For a better quantitative understanding, Fig. \ref{Figure5} shows the direct mapping $S_{q_1q_2}$ vs. $k_{zq_1q_2}$ for the same array configurations as in Fig. \ref{Figure4} and $|q_1|, |q_2|\le 50$. Apart from the trivial periodic case (Fig. \ref{Figure5}(a)), one observes highly-populated spectra, which display perfect periodicity in the case of commensurate scales (Fig. \ref{Figure5}(b)) and only some loose repetitiveness otherwise (Fig. \ref{Figure5}(c)). In both cases, a vast majority of the spatial frequencies have amplitude coefficients significantly smaller ($\lesssim -20$dB) than the dominant ones.

\subsection{Truncation Effects: Semi-Infinite Arrays}
\label{Semi-Infinite}
\begin{figure} 
\begin{center}
\includegraphics[width=10cm]{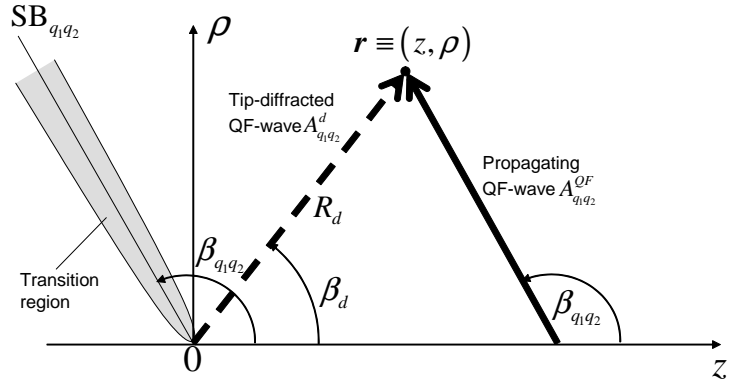} 
\end{center}
\caption{Quasi-Floquet (QF) wave phenomenologies for the infinite and semi-infinite array (planar cut through conical wavefields). For the infinite array, the wavefield away from the array axis is synthesized in terms of propagating QF waves $A^{QF}_{q_1q_2}$ (heavy solid arrow) arriving from direction $\beta_{q_1q_2}$ in (\ref{eq:betaq1q2}).
For the semi-infinite array, the region of validity of each propagating QF wave is limited by a conical shadow boundary SB$_{q_1q_2}$ with the same angle $\beta_{q_1q_2}$. A spherical tip-diffracted QF wave $A^d_{q_1q_2}$ (dashed arrow) arriving from direction $\beta_d$ ensures, via the transition function in (\ref{eq:UTD}), continuity of the wavefield across the parabolic transition region (gray shading) surrounding the shadow boundary cone.} 
\label{Figure3} 
\end{figure}

\begin{figure} 
\begin{center}
\includegraphics[width=12cm]{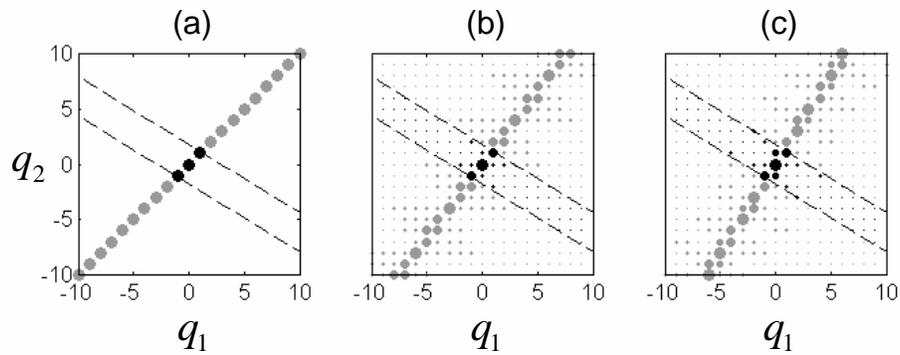} 
\end{center}
\caption{Modified-Fibonacci array with $d_{av}=1.1 \lambda_0, \eta=0$, and various values of the scale ratio $\nu=d_2/d_1$. ($q_1,q_2$)-plane mapping of propagating (dark dots) and evanescent (light dots) amplitude coefficients $S_{q_1q_2}$ in (\ref{eq:Wq}). Only $|q_1|,|q_2|\le 10$ spatial frequencies are considered, with dot size proportional to coefficient amplitude. The two oblique dashed lines delimit the propagating spectral range in (\ref{eq:prop}). (a): $\nu=1$ (periodic); (b): $\nu=0.75$; (c): $\nu=1/\tau$ (standard-Fibonacci).} 
\label{Figure4} 
\end{figure}

\begin{figure} 
\begin{center}
\includegraphics[width=10cm]{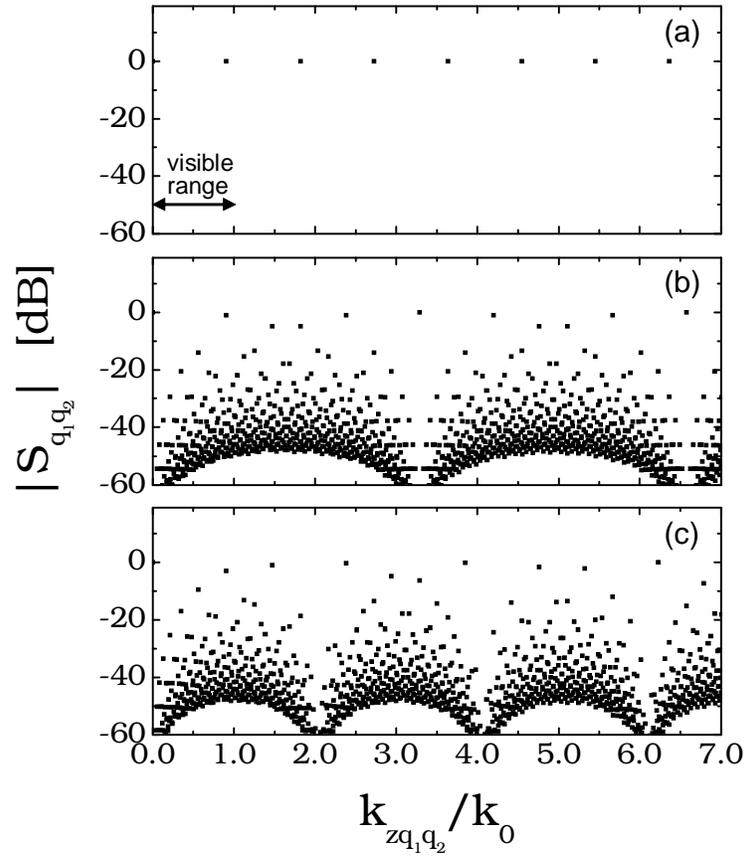} 
\end{center}
\caption{As in Fig. \ref{Figure4}, but $|S_{q_1q_2}|$ vs. $k_{zq_1q_2}$, for $|q_1|, |q_2|\le 50$ and $k_{zq_1q_2}\le7k_0$. Due to symmetry, only positive spatial frequencies are shown.} 
\label{Figure5} 
\end{figure}

Capitalizing on some analogies with the problem expounded in \cite{Capolino2}, we have also studied truncation-induced diffraction effects. For the semi-infinite ($m\ge 0$) version of the array in (\ref{eq:dist1}), one obtains a {\em truncated} QF wave synthesis,
\begin{eqnarray}
A({\bf r})&=&
\sum_{m=0}^{\infty}A_m({\bf r})\nonumber\\
&=&\sum_{m=0}^{\infty}\frac{\exp\left(-jk_0R_m\right)}{4\pi R_m} \exp\left(-jk_0\eta z_m\right)\nonumber\\
&=&\frac{A_0({\bf r})}{2}+\sum_{q_1,q_2=-\infty}^{\infty}S_{q_1q_2}A^T_{q_1q_2}({\bf r}),
\label{eq:TQFS}
\end{eqnarray}
where $A^T_{q_1q_2}$ are {\em truncated} QF wave propagators,
\begin{eqnarray}
A^T_{q_1q_2}({\bf r})&=&
\frac{1}{d_{av}}\int_{0}^{\infty}
\frac{\exp\left(-jk_0R\right)}{4\pi R}\exp\left(-jk_{zq_1q_2}z'\right)dz'\nonumber\\
&=&
\frac{\exp\left(-jk_{zq_1q_2}z\right)}{8 \pi d_{av}}
\int_{-\infty}^{\infty}
\frac{\mbox{H}_0^{(2)}\left(k_{\rho q_1 q_2}\rho\right)}{k_z-k_{z q_1 q_2}}
dk_z.
\label{eq:TQFW}
\end{eqnarray}
Exploiting the uniform high-frequency asymptotic approximation given in Eqs. (28)--(35) of \cite{Capolino2} for the spectral integral in (\ref{eq:TQFW}), one obtains
\beq
A^T_{q_1q_2}({\bf r})\sim
A^{QF}_{q_1q_2}({\bf r})U\left(\beta_{q_1q_2}^{SB}-\beta_d\right)+
A^d_{q_1q_2}({\bf r}).
\label{eq:UTQF}
\eeq
In (\ref{eq:UTQF}), $A^{QF}_{q_1q_2}$ is the QF wave propagator in (\ref{eq:PQFW}), $U$ denotes the Heaviside unit-step function, $\beta^{SB}_{q_1q_2}$ delimits the shadow-boundary of each QF wave (for {\em propagating} QF waves,  
$\beta_{q_1q_2}^{SB}=\beta_{q_1q_2}$). Moreover, $A^d_{q_1q_2}$ represents the {\em diffracted} QF wave emanating from the array tip,
\beq
A^d_{q_1q_2}({\bf r})\sim
\frac{\exp\left(-jk_0R_d\right)}{j4\pi d_{av}k_0R_d}
\frac{F(\gamma^2_{q_1q_2})}{\left(\cos\beta_{q_1q_2}-\cos\beta_d\right)},
\label{eq:DQF}
\eeq
where $R_d=\sqrt{\rho^2+z^2}$, $\beta_d=\arctan(\rho/z)$, $\gamma_{q_1q_2}=\sqrt{2 k_0 R_d}\sin[(\beta_{q_1q_2}-\beta_d)/2]$, and $F$ denotes the standard transition function of the uniform theory of diffraction (UTD),
\begin{eqnarray}
F(x)=2j\sqrt{x}\exp(jx)\int_{\sqrt{x}}^{\infty}\exp\left(-j\xi^2\right)d\xi,\nonumber\\
-\frac{3}{2}\pi<\mbox{arg}(x)\le\frac{\pi}{2}.
\label{eq:UTD}
\end{eqnarray}
The wave phenomenologies are illustrated in Fig. \ref{Figure3}, for the case of {\em propagating} ($|k_{zq_1q_2}|<k_0$) QF waves. The region of validity of each QF wave $A^{QF}_{q_1q_2}$ (heavy arrow in Fig. \ref{Figure3}) arriving from direction $\beta_{q_1q_2}$ is now limited by a conical shadow boundary. The spherical tip-diffracted QF wave $A^d_{q_1q_2}$ (dashed arrow in Fig. \ref{Figure3}) arriving from direction $\beta_d$ ensures, via the transition function in (\ref{eq:UTD}), continuity of the wavefield across the parabolic transition region (gray shading in Fig. \ref{Figure3}) surrounding the shadow boundary cone.

Note that {\em evanescent} ($|k_{zq_1q_2}|>k_0$) QF waves yield negligible contributions at observation points far from the array axis, but they excite {\em propagating diffracted} fields that need to be taken into account. As in \cite{Capolino2} (see the discussion after Eq. (35) there), these contributions are approximated via {\em nonuniform} asymptotics ($F\approx 1$ in (\ref{eq:DQF})). We point out that, at variance with the periodic array case (cf. Eq. (36) in \cite{Capolino2}), it is not possible here to recast the total spherical wave diffracted field in a more manageable form.

\section{Results and Potential Applications}
\label{Res_App}
Although our main interest in this preliminary investigation is focused on wave-dynamical phenomenologies associated with radiation from quasiperiodic antenna arrays, we briefly discuss some computational and applicational aspects with the hope of providing further insights.
We stress that no attempt has been made at this stage to devise {\em optimal} computational schemes, nor do we deal with actual fabrication-oriented issues (feeding, matching, inter-element coupling, etc.).  

\subsection{Numerical Results}
\label{Results}
\begin{figure} 
\begin{center}
\includegraphics[width=10cm]{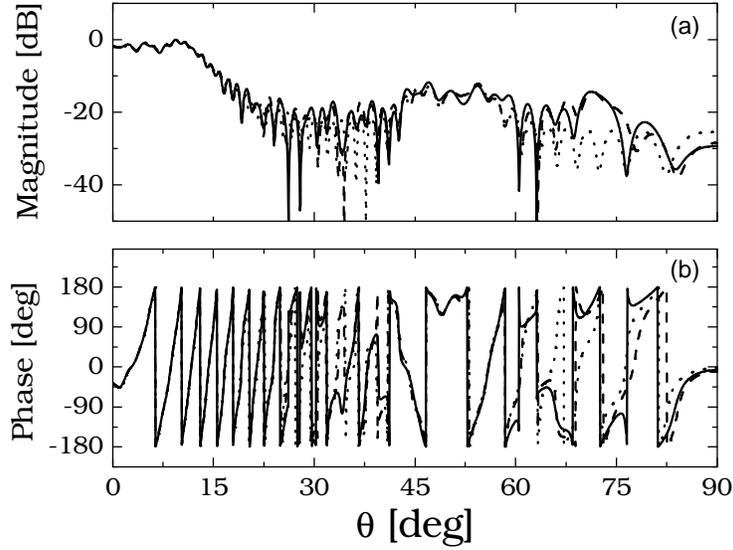} 
\end{center}
\caption{$101$-element standard-Fibonacci ($\nu=1/\tau$) array with $d_{av}=0.5\lambda_0$ and $\eta=0$. Near-zone ($R=100\lambda_0$) normalized potential field scan. {\bf -----} Reference solution (element-by-element summation); {\bf $\cdots \cdots$} QF synthesis retaining $N_p=3$ dominant propagating waves (selected within $|q_1|, |q_2|\le 50$); {\bf - - -} QF synthesis retaining $N_p=9$ dominant propagating QF waves. Due to symmetry, only positive angles are shown.} 
\label{Figure6} 
\end{figure}

\begin{figure} 
\begin{center}
\includegraphics[width=10cm]{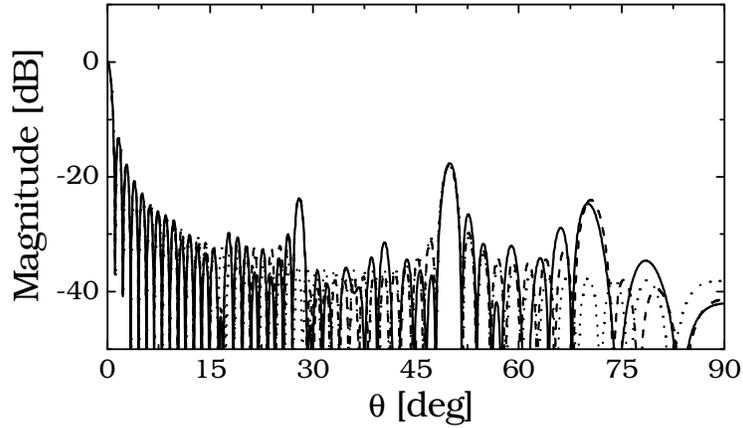} 
\end{center}
\caption{As in Fig. \ref{Figure6}, but far-field pattern.} 
\label{Figure7} 
\end{figure}

\begin{figure} 
\begin{center}
\includegraphics[width=10cm]{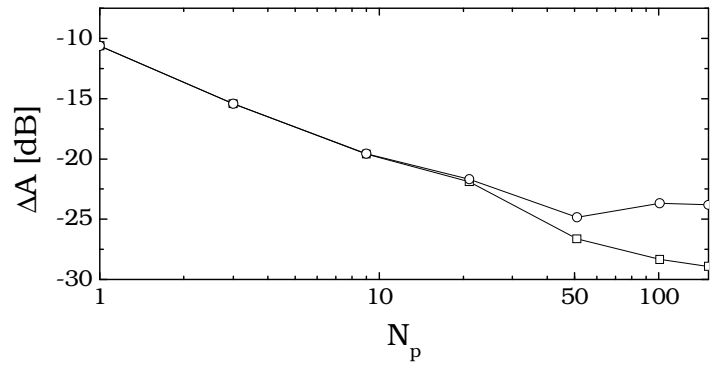} 
\end{center}
\caption{Parameters as in Fig. \ref{Figure6}, but r.m.s. error $\Delta A$ in (\ref{eq:rms}) vs. number $N_p$ of propagating waves retained in the near field ($R=100\lambda_0$) QF synthesis. 
Circular bullets: No evanescent diffracted waves retained. Square bullets: $N_e=N_p$ evanescent diffracted waves retained. Results for the far-field are practically identical.} 
\label{Figure8} 
\end{figure}

\begin{figure} 
\begin{center}
\includegraphics[width=10cm]{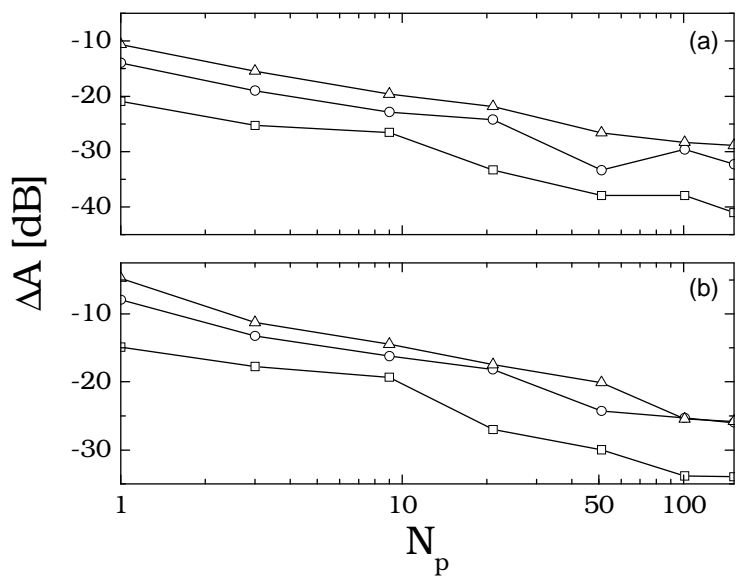} 
\end{center}
\caption{As in Fig. \ref{Figure8}, with $R=100\lambda_0$, $N_e=N_p$, and various values of scale ratio $\nu=d_2/d_1$ and average spacing $d_{av}$. (a): $d_{av}=0.5\lambda_0$. (b): $d_{av}=0.75\lambda_0$.
Square bullets: $\nu=0.9$. Circular bullets: $\nu=0.75$. Triangular bullets: $\nu=1/\tau$.} 
\label{Figure9} 
\end{figure}

From a computational viewpoint, the actual utility of the QF syntheses in Sec. \ref{Radiation} ({\em double} summations involving an {\em infinite} number of propagating waves) might appear questionable as compared with brute-force element-by-element synthesis. However, as observed in Sec. \ref{Infinite} (see also Fig. \ref{Figure5}), under appropriate conditions, a large number of propagating QF waves could be {\em weakly} excited, thus suggesting the possibility, yet to be explored, of devising effective truncation schemes.
Here, we try to quantify some of these aspects via numerical examples. We begin by considering a 101-element nonphased standard-Fibonacci array ($\eta=0$, $\nu=1/\tau$, $|m|\le 50$ in (\ref{eq:dist1})). The truncated QF synthesis developed in Sec. \ref{Semi-Infinite} for a semi-infinite array can readily be exploited here by expressing the finite array interval as the difference between two overlapping semi-infinite intervals. Figure \ref{Figure6} shows the near-field ($R=100\lambda_0$) QF synthesis results for $d_{av}=0.5 \lambda_0$, using the crudest possible truncation criterion based on retaining $N_p$ dominant propagating QF waves in (\ref{eq:TQFS}). Also shown, as a reference solution, is the element-by-element synthesis in (\ref{eq:elbyel}) (with $|m|\le 50$). It is observed that the QF synthesis with $N_p=9$ propagating waves provides reasonably good agreement, and even $N_p=3$ is still capable of fleshing out most of the wavefield structure. Similar statements can be made for the far-field pattern in Fig. \ref{Figure7}. In order to better quantify the accuracy, and address convergence issues, we have computed the r.m.s. error
\beq
\Delta A(R)=\sqrt{\frac{\int_{-\pi/2}^{\pi/2}\left|A^{RS}(R,\theta)-A^{QF}(R,\theta)\right|^2 d\theta}
{\int_{-\pi/2}^{\pi/2}\left|A^{RS}(R,\theta)\right|^2d\theta}},
\label{eq:rms}
\eeq
where the subscripts ``$^{RS}$'' and ``$^{QF}$'' denote the reference solution and the QF synthesis, respectively.
Figure \ref{Figure8} shows the error behavior vs. the number $N_p$ of dominant propagating QF waves retained, for near-field synthesis (results for the far-field are practically identical). Also shown, for comparison, are results obtained by retaining a number $N_e=N_p$ of dominant evanescent QF diffracted waves. It is observed that a moderate number ($\sim 10$) of QF waves is capable of providing acceptable accuracy ($\Delta A\sim -20$dB). The contribution of $N_e$ retained evanescent QF diffracted waves is practically negligible for $N_e\lesssim 20$, but can give observable improvements at larger $N_e$-values.
Intuitively, one would expect the convergence behavior to improve in the presence of weaker aperiodicity ($\nu\approx1$) and smaller average inter-element spacing, and viceversa. This is confirmed by the results shown in Fig. \ref{Figure9}, for two values of the inter-element spacing ($d_{av}=0.5\lambda_0$ and $0.75\lambda_0$) and three values of the scale ratio ($\nu=0.9, 0.75$ and $1/\tau$).

To sum up, the above results seem to indicate that moderate-size QF syntheses, truncated by using even crude criteria, can still be capable of capturing the dominant features of the relevant wave dynamics. However, besides the computational convenience, which can become questionable if highly accurate results are needed, the QF parameterization can offer valuable insights for judicious exploitation of the inherent degree of freedom (scale ratio $\nu=d_2/d_1$) in the array (see, e.g., Sec. \ref{Applications}).

\subsection{Potential Applications}
\label{Applications}
In view the {\em discrete} character of its spectrum (see (\ref{eq:Fib_spect2})), the modified-Fibonacci array in (\ref{eq:dist1}) does not offer particular advantages within the framework of array thinning, as compared with periodic arrays. In this connection, other 1-D aperiodic sequences, such as the Rudin-Shapiro \cite{Vasco} (characterized by {\em continuous} spectra), might be worth being explored.

\begin{figure} 
\begin{center}
\includegraphics[width=10cm]{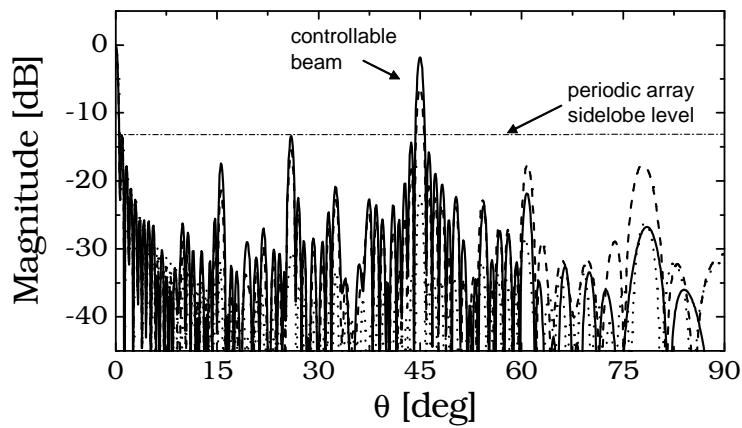} 
\end{center}
\caption{$101$-element modified-Fibonacci array with $d_{av}=0.874\lambda_0$ and $\eta=0$. 
Radiation pattern (array factor) $|{\hat F}(k_0\sin\theta)|$ in (\ref{eq:Fib_spect1}) (with $|m|\le 50$) for various values of scale ratio $\nu$, illustrating the controllable multibeam capability (see Sec. \ref{Applications}). {\bf -----} $\nu=0.25$ ($S_{01}=-1.83$dB);  {\bf - - -} $\nu=0.5$ ($S_{01}=-6.37$dB); 
{\bf $\cdots \cdots$} $\nu=0.9$ ($S_{01}=-23.3$dB). Note that minor sidelobes never exceed the  periodic-array sidelobe level (-13dB).}
\label{Figure10} 
\end{figure}

However, the degree of freedom available in the choice of the scale ratio $\nu=d_2/d_1$ can be exploited in principle to control the spectral properties and achieve, for instance, a {\em multibeam} radiation pattern. As a simple example, we consider a nonphased ($\eta=0$) configuration with average inter-element spacing $d_{av}<\lambda_0$. It is readily observed from (\ref{eq:Wq}), (\ref{eq:spat_freq}) that, besides the unit-amplitude QF wave ($q_1=q_2=0$) (main beam at broadside), the only other propagating QF waves (secondary beams) are
\begin{subequations}
\beq
q_1=-q_2=\pm1,~~|W_{q_1q_2}|=\frac{\pi(1+\tau)(1+\nu)}{\nu+\tau}>\pi,
\label{eq:beam1}
\eeq
\beq
q_1=\pm1, q_2=0,~~|W_{q_1q_2}|=\frac{\pi(1+\tau)}{\nu+\tau}\ge\pi,
\label{eq:beam2}
\eeq
\beq
q_1=0, q_2=\pm1,~~|W_{q_1q_2}|=\frac{\pi\nu(1+\tau)}{\nu+\tau}\le\pi.
\eeq
\label{eq:beam3}
\end{subequations}
Note that, since $|W_{q_1q_2}|\ge \pi$, the amplitude of the secondary beams in (\ref{eq:beam1}) and (\ref{eq:beam2}) will {\em always} be at least 13dB below the main beam, and thus at the same level of sidelobes as in finite-size periodic arrays, {\em irrespective} of the scale ratio $\nu$. Conversely, for the secondary beam in (\ref{eq:beam3}), one has $|W_{q_1q_2}|\le\pi$, and thus its amplitude can be controlled over a wide range by varying the scale ratio $\nu$. Restricting attention to the ($q_1=0, q_2=1$) beam, one finds from (\ref{eq:Wq}), (\ref{eq:spat_freq}) the corresponding direction $\beta_{01}$ (from endfire) and amplitude $S_{01}$ to be
\beq
\beta_{01}=\arccos\left[\frac{\lambda_0 \tau}{d_{av}(1+\tau)}\right],~~~S_{01}=\frac{({\nu+\tau})\sin\left[\frac{\pi\nu(1+\tau)}{\nu+\tau}\right]}{\pi\nu(1+\tau)}.
\label{eq:beta01}
\eeq
The direction $\beta_{01}$ can thus be steered, by varying $d_{av}/\lambda_0$, up to a maximum value
$\beta_{01}^{(max)}=\arccos\left[\tau/(1+\tau)\right]\approx51.83^o$, corresponding to the maximum spacing ($d_{av}=\lambda_0$) allowable to prevent emergence, in the visible range, of higher-order grating lobes.
The amplitude $S_{01}$ can be controlled (from 0 to 1, in principle) by varying the scale ratio $\nu=d_2/d_1$ (from 1 to 0). One thus obtains a {\em multibeam} radiation pattern with the possibility of controlling the secondary beam amplitude. Results for the ($q_1=0, q_2=-1$) beam follow from symmetry considerations.
Obvious fabrication-related issues prevent $\nu$ from being exceedingly small, but values of $\nu\approx0.25$ are sufficient to achieve secondary beam amplitudes $\sim -2$dB. Figure \ref{Figure10} shows the radiation pattern (array factor) of a 101-element array, with $d_{av}/\lambda_0$ and $\nu$ chosen from (\ref{eq:beta01}) so as to create a secondary beam at $45^o$ with various amplitudes. It is observed that actual amplitude values are very close to the infinite-array predictions in (\ref{eq:beta01}), and that minor sidelobes never exceed the periodic-array sidelobe level (-13dB).

It is hoped that the above observations might open up new perspectives for reconfigurable arrays. Note that {\em rational} values of $\nu$ (commensurate scales) correspond to configurations interpretable as {\em periodic} arrays with {\em aperiodically-distributed} lacunas. In these cases, one can think of easily-implementable reconfigurable strategies, based on switching from {\em single-beam}  (periodic array) to {\em multibeam} (modified-Fibonacci) radiation patterns, via on-off selection of a set of antenna elements, keeping the average inter-element spacing $d_{av}<\lambda_0$.

\section{Conclusions and Perspectives}
\label{Conclusions}
A simple illustrative example of wave interaction with quasiperiodic order has been discussed in connection with
the radiation properties of 1-D antenna arrays, utilizing the modified-Fibonacci sequence. A ``quasi-Floquet'' analytic parameterization, based on a generalized Poisson summation formula, has been derived for infinite and semi-infinite arrays. Computational issues and potential applications have been briefly addressed.

It is hoped that the prototype study in this paper, through its instructive insights into some of the basic mechanisms governing wave interactions with quasiperiodic order, may lead to new applications in array radiation pattern control,
in view of the additional degrees of freedom available in aperiodic structures.
Accordingly, we are planning current and future investigations of modified-Fibonacci arrays that will emphasize the effects of the scale ratio parameter ($\nu=d_2/d_1$) on the input impedance as well as on the coupling to possible dielectric-substrate-induced leaky modes. Exploration of the radiation properties of antenna arrays based on other well-understood aperiodic sequences (e.g., Thue-Morse, period-doubling, Rudin-Shapiro \cite{Vasco}) is also being pursued.

\section*{Appendix\\Pertaining to (\ref{eq:Fib_spect})--(\ref{eq:Gen_Poiss})}
In \cite{Buczek}, the modified-Fibonacci array spectrum in (\ref{eq:Fib_spect}) is computed using two equivalent approaches, one directly based on the cut-and-project scheme described in Sec. \ref{Mod_Fib} (see also Fig. \ref{Figure2}), and the other based on an ``average unit cell'' method \cite{Wolny}. Although the two approaches are relatively simple in principle, their implementation is rather involved and not reported here for brevity. The reader is referred to \cite{Buczek}, \cite{Wolny} for theoretical foundations and technical details, and to \cite{Kolar1}, \cite{Kolar2} for an alternative approach applicable to rather general substitutional sequences.
We point out that in \cite{Buczek}, the result for the radiation spectrum is given in normalized form, with a non-explicit multiplicative constant, and assuming zero phasing ($\eta=0$). In the paper here, the possible presence of phasing is accounted for via the spectral shift $k_0\eta$ in (\ref{eq:spat_freq}). The calculation of the proper multiplicative constant in (\ref{eq:Fib_spect}) (and hence that in (\ref{eq:Gen_Poiss})) has been accomplished by first observing that for the modified-Fibonacci sequence in (\ref{eq:ModFib}), the number of array elements falling within a window of width $2\zeta$ approaches the ratio between the window width and the average inter-element spacing $d_{av}$ in (\ref{eq:dav}), as $\zeta\rightarrow\infty$, i.e.,
\beq
\sum_{m=-\infty}^{\infty}\left[U\left(z_m+\zeta\right)-U\left(z_m-\zeta\right)\right]\sim \frac{2\zeta}{d_{av}},
\label{eq:app1}
\eeq
with $U$ denoting the Heaviside step-function. It then follows from (\ref{eq:dist1}), assuming $\eta=0$ and recalling (\ref{eq:app1}), that
\begin{eqnarray}
&&\lim_{\zeta \rightarrow \infty}\frac{1}{2\zeta}
\int_{-\zeta}^{\zeta}f(z)dz\nonumber\\
&=&\lim_{\zeta \rightarrow \infty}\frac{1}{2\zeta}
\sum_{m=-\infty}^{\infty}\left[U\left(z_m+\zeta\right)-U\left(z_m-\zeta\right)\right]\nonumber\\
&=&\lim_{\zeta \rightarrow \infty}
\frac{1}{2\zeta}\left(\frac{2\zeta}{d_{av}}\right)=\frac{1}{d_{av}}.
\label{eq:mult}
\end{eqnarray}

The proper multiplicative constants in (\ref{eq:Fib_spect}) and (\ref{eq:Gen_Poiss}) have accordingly been computed by requiring that (\ref{eq:Gen_Poiss}) satisfy (\ref{eq:mult}) for $\eta=0$.

\section*{Acknowledgements}
L.B. Felsen acknowledges partial support from Polytechnic University, Brooklyn, NY 11201, USA.


\newpage
%

%

%

%

%

%

%

%

%

%

\end{document}